# Are Computer Science and Engineering Graduates Ready for the Software Industry? Experiences from an Industrial Student Training Program

Eray Tuzun
Bilkent University
Department of Computer Engineering
*and*
Havelsan
Technology and Academy Directorate
Turkey
eraytuzun@cs.bilkent.edu.tr

Hakan Erdogmus
Carnegie Mellon University
Electrical and Computer Engineering
USA
hakan.erdogmus@sv.cmu.edu

Izzet Gokhan Ozbilgin
Havelsan
Technology and Academy Directorate
Turkey
gozbilgin@havelsan.com.tr

## ABSTRACT

It has been 50 years since the term "software engineering" was coined in 1968 at a NATO conference. The field should be relatively mature by now, with most established universities covering core software engineering topics in their Computer Science programs and others offering specialized degrees. However, still many practitioners lament a lack of skills in new software engineering hires. With the growing demand for software engineers from the industry, this apparent gap becomes more and more pronounced. One corporate strategy to address this gap is for the industry to develop supplementary training programs before the hiring process, which could also help companies screen viable candidates. In this paper, we report on our experiences and lessons learned in conducting a summer school program aimed at screening new graduates, introducing them to core skills relevant to the organization and industry, and assessing their attitudes toward mastering those skills before the hiring process begins. Our experience suggests that such initiatives can be mutually beneficial for new hires and companies alike. We support this insight with pre- and post-training data collected from the participants during the first edition of the summer school and a follow-up questionnaire conducted after a year with the participants, 50% of whom were hired by the company shortly after the summer school.

## 1 INTRODUCTION

Havelsan [1] is a medium-size Turkish software and systems company operating in the simulation, defense and IT sectors. It was founded three decades ago. In the last decade, the company has rapidly grown to more than 1,400 employees. Software development is the major activity in the company, so the majority of the engineers have computer science, software engineering, or computing-related engineering degrees. In 2016, the company created a corporate training department, Havelsan Academy, to support its employees' vocational and technical development needs. The department is also responsible for industry-academia relations and academic outreach to attract top talent.

One of the first major initiatives of Havelsan Academy was the institution of the Havelsan Summer School (hereby the *summer school*). Now well-established, the initiative has three main objectives: (a) hiring top-notch new graduates; (b) increasing the visibility of the company within universities; (c) giving back to the academic community by supporting their goal of educating highly-qualified personnel.

Under goal (a), as an important side benefit, the initiative also aims at addressing the gap between the new hires' skill levels and the company's expectations of them. This gap, well-acknowledged in the industry, has been studied and highlighted by several researchers [2-4].

The gap can potentially be tackled in several ways: by updating the curriculum via transfer of experience from industry to academia [5]; via training as part of post-employment onboarding of new hires [6]; and by partially relying on certification programs, such as those provided by IEEE [7]. However, Havelsan Academy decided to take a different approach: creation of a summer school to replace post-employment training with pre-employment training and assessment. Notably, his approach flips the normal hiring process on its head by starting the onboarding process early, even before an employee becomes a potential hire. This paper details the experience and insights gained from the first edition of the summer school over a year after its founding to reflect on how it met its original goals and what was learned during the process to improve its future offerings.

The remainder of the paper is organized as follows. Section 2 reviews the related work. Section 3 presents summer school setup and process while Section 4 describes its execution. Section 5 presents the central results and observations. Section 6 discusses the benefits, economic considerations, lessons learned, and limitations. Finally, Section 7 sums up our conclusions, with an emphasis on the organizational impact of the summer school program.

## 2 RELATED WORK

Various studies have explored the knowledge gap between software education and software industry's expectations of new hires [8]–[10]. Based on a large survey of software professionals, Lethbridge [9] concluded that "certain software engineering topics should be given more emphasis, while the emphasis on certain mathematics topics should be changed." These results pushed the author's university to align its software engineering curriculum to cover the top topics emphasized by professionals. Kitchenham et al. [11] adopted Lethbridge's survey instrument in the UK to compare selected UK curricula to industry expectations, and reported similar conclusions, with the additional caveat that the set of topics were taught far less than their importance would suggest. Radermacher and Walia's systematic review [12] confirms these finding over a decade after Lethbridge's study, the gaps largely remaining the same. The study by Aasheim et al. [8], conducted with hundreds of IT managers, emphasized the relevance of work experience over academic performance measures such as GPA, and suggested curricular revisions to address the discrepancy between what is valued in academia and in industry. All of these approaches attempt to better align academic curricula with industrial expectations to bridge the skill gap.

We have not been able to find any published reports of rigorous initiatives by companies addressing the gap through systematic, pre-employment training targeted at graduating students. Summer internships, co-op terms, and mentoring programs such as Google Summer of Code are a form of gap-related training, but we make a distinction between *on-the-job* initiatives before full-time employment and curricula-based approaches targeting specific knowledge areas. It is possible that the initiatives of the latter kind exist or have existed, although they have not been reported publicly.

There has however been work on the transfer of knowledge and pedagogical techniques from the industry to a university setting. Bleek et al. [5], for example, discuss their experiences with transferring the teaching methodology used in the industry to software engineering curricula at the University of Hamburg. The authors, who have both industrial and academic experience, point out fundamental differences in teaching methodology between the two settings, which they addressed by increasing hands-on and lab components, intensity of feedback, one-on-one interactions with the instructors, and project-based learning. Many modern academic software engineering courses in fact adopt such techniques [13], however these strategies are still embedded in the academic system, and hence cannot take into account organizational context and specific needs addressable best through programs administered inside the company.

Outside academia and industry, third-party software engineering certification programs offered by professional societies and governments also attempt to fill in the skills gap. These programs promise to level the playing field for new graduates through testing standardized knowledge buckets, coupled with mentored internship. For example, in the US, the Texas state government has been certifying professional software engineers since 1999 [14]. IEEE's Professional Competency Certifications program for software developers is another popular example [7].

As for post-employment strategies, systematic, within-company programs are not uncommon, especially in large companies, although not always publicized. One such large scale initiative was undertaken by Siemens to train over 500 company engineers over three years. Samarthyam et al. [6] describe this program, whose curriculum was based on IEEE's Software Engineering Body of Knowledge [15].

## 3 SUMMER SCHOOL SETUP

The process followed to set up and execute the summer school is shown in Figure 1. The process consisted of a preparation phase, execution of classes with exams (the column "Classes"), and a follow-up survey. The first and last day of classes had special initiation and close-up activities, so they are shown in separate columns. Preparations started in April 2016, and the last phase, the follow-up survey, was completed in October 2017. In the following subsections, we elaborate on the central parts of the preparation process.

### 3.1 Curriculum Setup

To setup the curriculum, we relied on two curricular analyses of Computer Science, Computer Engineering, and Software Engineering programs in Turkey [16, 17]. We used these analyses to identify both foundational areas and gaps between academic education and industrial practice. Our main goal in forming the curriculum was to focus on knowledge areas that were not sufficiently emphasized in university settings.

It was not possible to address all gaps, and in the end, we settled on eight knowledge areas of high priority: *Software Verification and Validation*, *System and Software Architecture*, *Agile Software Development*, *Software Project Management*, *Application Lifecycle Management*, *JavaScript Programming and Web Development*, *Object-Oriented Programming with Java*, and *Software Craftsmanship*. Each of these areas became a separate class with its own specialized syllabus and instructor.

We also decided to add a *Software Career Panel* as a separate module to give students a better appreciation of the different software industry roles that exist in the company, such as requirements engineer, architect, development engineer, test engineer, and project manager.

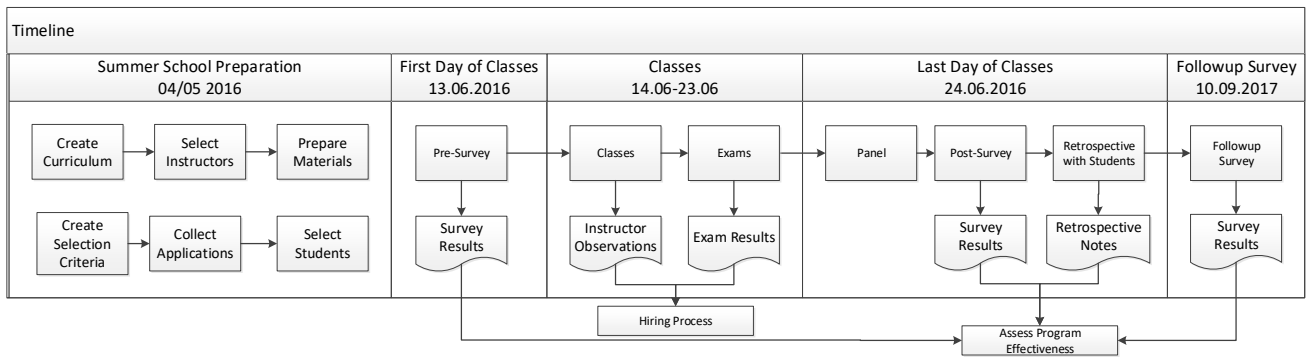

**Figure 1:** Summer school process and timeline.

## 3.2 Instructor Selection and Alignment

Once we identified the knowledge areas and associated classes, we combed the organization for potential instructors and approached them. To identify the candidates, for each class in the curriculum, we looked for the following two qualities in instructors:

1. subject matter experience demonstrated on the job, and
2. previous teaching experience either inside the company (internal training programs) or more formal teaching experience outside the company in a higher-education setting.

We performed multiple meetings with potential instructors to agree on goals, processes, how to setup the syllabi, and how to keep the curriculum coherent within classes and consistent across classes. For some of the classes, we selected two instructors to balance the load and sufficiently cover the expertise required.

The average industry experience of the instructors was 13 years. Of the 10 instructors selected, two had a PhD degree and four had a Master's degree. The rest of them had a Bachelor's degree.

Once we had the instructor team on board, we collectively settled on an exam strategy and came up with consistent exam criteria for all of the classes.

## 3.3 Defining Learning Outcomes

For each class in the curriculum, we asked the assigned instructors to come up with three to five learning outcomes for their class to guide the class syllabi. The summer school organizer, who also served as the instructor team lead, reviewed the learning outcomes, and gave feedback to the other instructors. The learning outcomes were revised and finalized following this feedback.

For example, the following learning outcomes were defined for the *Agile Software Development* class:

- List principles and practices used in agile software development;
- Grasp the elements of Agile Manifesto;
- Identify the concepts of Scrum (roles, practices, and artifacts);
- Develop the ability to function in a Scrum Team.

## 3.4 Student Recruitment

To select the most qualified students, we announced the event nationwide on a website and via a targeted campaign. We prepared a poster and sent it to all major Turkish universities. We also promoted the event on main social media platforms.

Applications were accepted online, subject to the following eligibility requirements:

- be a 3$^{rd}$- year or 4$^{th}$-year undergraduate student;
- pursuing an engineering or a computer science degree;
- have a minimum GPA of 3.0;
- have a good command of English; and
- have ability to attend classes full-time during the specified period.

The application form required the students to provide their education information, résumé, transcripts, and a short essay explaining their interest in the summer school.

We received a total of 170 valid applicants from 37 different universities and 18 different cities in Turkey. The selection criteria consisted of academic performance (GPA), relevant industry or project experience (internships), breadth of technical skills (programming languages, platforms, tools), and diversity (gender and geographical distribution). After careful screening, we picked 16 students to invite, with an admission rate of below 10%. Of the 16 invited students, 14 were 4$^{th}$-year and two were 3$^{rd}$-year students. Their departments/majors were: Computer Science/Engineering (7), Software Engineering (4), Electrical Engineering (4), and Industrial Engineering (1). All of the invited students attended and completed the summer school.

## 3.5 Miscellaneous Preparations

*Rooms and labs.* In the company, we have a dedicated building where internal training programs take place. This building has classrooms with a U-shaped, conference setup that can accommodate up to 18 trainees. We decided to use these rooms to conduct the theory portions of the classes. The company also has two computer labs for training purposes. We decided to use these labs for topics requiring hands-on work. For lab-based classes, the lab computers were set up with all the necessary software and disk images before the start date.

*Class syllabi.* Each instructor was asked to prepare a single-page description of the class, including a summary of the syllabus.

*Hand-outs.* The instructors were asked to prepare presentation slide decks in advance, with printed copies ready to be handed out to the students.

*Exams.* Each instructor was asked to prepare a one-hour exam to be conducted shortly after the completion of their class. Exams were scheduled in advance.

*Survey instruments, evaluation forms, and post-mortem strategy.* We carefully prepared pre-event and post-event surveys to deploy to the students upon entry into the program (on first day of classes) and exit from the program (on last day of classes). An instructor evaluation survey, to be completed by the students, was also developed. We also decided in advance on the format of the planned retrospective session with the students to be held at the end of the summer school.

## 4 SUMMER SCHOOL EXECUTION

The summer school kicked off with an opening ceremony, a company tour/orientation, and a student social on June 13, 2016. It concluded with a closing ceremony on June 24. In the following subsections, we describe the execution in detail.

### 4.1 Pre-Event Survey

Before the beginning of the classes, the students completed a pre-event survey. The students were informed that (1) the pre-event and post-event surveys and their instructor evaluations would not affect their performance assessment, and (2) their responses would be treated as anonymous and confidential. The pre-event survey consisted of four sections:

*Experience and Background:* To establish a baseline, we asked the students about their industry experience (including internships and part-time work), knowledge of programming languages and software technologies, and previous software-related classes they had taken.

*Aspirations:* Open-ended questions in this section first probed the students' awareness of various software engineering disciplines, followed by their career aspirations about different software engineering roles and jobs. We also asked about their short-term plans and expectations from the summer school.

*Familiarity with and Appreciation of Software Engineering Concepts:* This section gauged the students' breadth of knowledge and understanding of software engineering. We first asked them to list the activities that they thought a professional software engineer would perform on a day-to-day basis on the job. We then gave them a long list of software engineering concepts that cover a wide scope of topics (ranging from technical practices and requirements techniques to design principles and management concepts) and asked the students to rate their familiarity with each. Finally, based on the summer school curriculum, we asked them to rate their familiarity with a few specific topics covered in each class. Some example questions from this section are shown in Figure 2 for the *Application Lifecycle Management* class.

### 4.3 Post-Event and Evaluation Surveys

A post-event survey was administered on the final day of the class after the closing ceremony. The post-event survey repeated the questions asked in the pre-event survey on the students' career aspirations and their familiarity and appreciation of software engineering concepts. The goal of the post-event survey was to assess any shifts in student responses relative to the pre-event survey as a result of the students' exposure to the summer school.

1. **What is your experience level with software development processes or methodologies?**

○ I am not familiar with software development processes.

○ I am familiar with a few software development processes, with either traditional or modern.

○ I am familiar with a few software development processes, including both traditional and modern

○ I am knowledgeable about software development processes and I understand the advantages and disadvantages of different schools of thoughts.

○ I am knowledgeable about software development processes, understand their advantages and disadvantages, and worked with both traditional and modern processes.

☐ *Check this box if you have taken a course on software development processes or covered the subject in class.*

2. **How knowledgeable are you with the different software development roles that exist in a software development team?**

○ I do not know about the different roles in a software development team.

○ I am familiar with a few basic roles in a software development team.

○ I am familiar with most central roles in a software development team.

○ I have a fairly good understanding of the different roles, along with the associated responsibilities, importance, and interactions.

○ I have a fairly good understanding of the different roles in a software development team and have played more than one role in a real-world context.

**Figure 2: Example questions from the pre-event survey.**

In addition to the post-event survey, the students filled out a multi-part evaluation survey on the organization of the event and the performance of the instructors. Organization-related questions asked about the application and registration experience, classroom conditions, and amenities provided.

**Table 1: Summer School Schedule**

| Day | Topics Covered |
|---|---|
| 1 | Opening Ceremony; Orientation; Pre-event Survey; Application Lifecycle Management (C1) |
| 2 | C1 cont'd; System & Software Architecture (C2) |
| 3 | C1 Exam; C2 cont'd; Object-Oriented Programming with Java (C3) |
| 4 | C2 Exam; C3 cont'd; Software Craftsmanship (C4) |
| 5 | C3 Exam; C4 cont'd; JavaScript Programming and Web Development (C5) |
| 6 | C4 Exam; Software Verification and Validation (C6) |
| 7 | C6 Exam; C5 cont'd |
| 8 | C5 Exam; Agile Software Development (C7) |
| 9 | C7 Exam; Software Project Management (C8) |
| 10 | C8 Exam; Software Engineering Career Panel; Instructor Evaluation Survey; Post-Event Survey; Retrospective; Graduation and Closing Ceremony |

## 4.2 Execution of Classes

Classes were conducted on a dense schedule every day from 9h00 to 16h30 in blocks of 50-minute lectures/labs, each block followed by a 10-minute break or lunch. A daily 90-minute lunch break was held to allow students time to gel, network, and socialize. Exams started on the third day and were held every morning from 8h00 to 9h00. Table 1 shows the daily schedule.

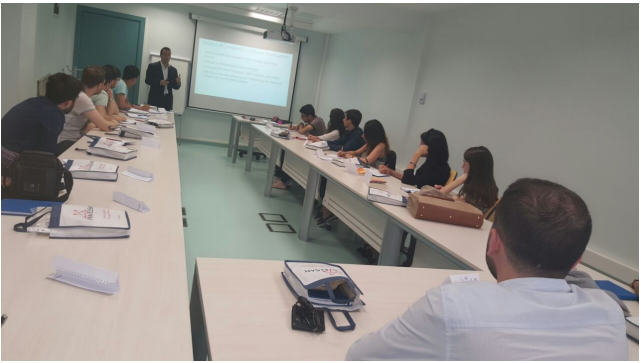

**Figure 3: Lecture in a U-shaped classroom.**

For the classes on *Agile Software Development*, *JavaScript Programming and Web Development*, *Object-Oriented Programming*, and *Software Craftsmanship*, the instructors preferred to use the lab format shown in Figure 4. The rest of the classes were conducted in a conference-room setting, shown in Figure 3, to encourage interactivity.

The instructor evaluation survey contained questions related to both course content and delivery. Course content questions covered clarity, sufficiency, appropriateness, general structure, logical sequence, integrity, and quality of instructional materials. Delivery questions focused on importance of the subject matter, instructors' ability to communicate the subject matter, and effectiveness and use of examples. Finally, open-ended questions solicited suggestions for improvement.

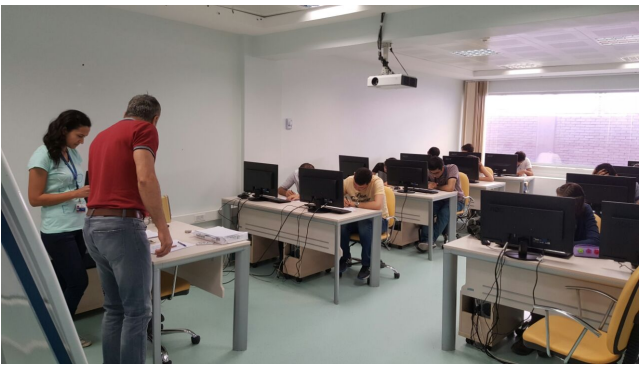

**Figure 4: Scene from a lab session.**

## 4.4 Retrospective

After the graduation ceremony, we conducted a post-mortem activity with the students to obtain more information on how the students felt about the summer school and how to improve future editions. We used the *Mad-Sad-Glad* retrospective format (https://www.retrium.com/resources/techniques/mad-sad-glad), a post-mortem technique popular in the agile software development community. Before the retrospective start, we divided a whiteboard into three sections labeled *Mad* (aspects that spoiled the whole fun), *Sad* (aspects that needed improvement), and *Glad* (aspects that worked well and should be retained).

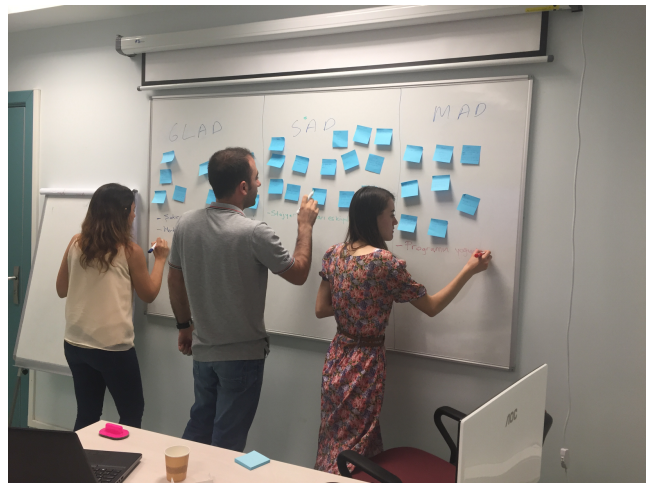

**Figure 5: Retrospective session.**

After explaining the technique, we divided the students into four equally-sized groups and equipped them with stacks of sticky notes. We asked the groups to first spend 15 minutes on individual ideation, and then 30 minutes as a group on brainstorming the individual ideas to populate the *Mad*, *Sad*, and *Glad* categories (Figure 5). After this part, we asked all groups to collectively organize their sticky notes into logical themes to create an affinity map. In the final round, we performed *dot-voting* to identify the most important themes in each category. Finally, we discussed in a plenary session the top-voted themes to solicit clarification and concrete improvement suggestions. The moderators took notes.

## 4.4 Follow-up Survey

Two months after the first anniversary of the summer school, we conducted an online follow-up survey with the participants to assess if they had been able to use and apply the learnings from the summer school. In the survey, we asked the respondents about (a) their current positions and (b) the benefits of their summer school experience in their job search and in the first year of their careers as a software professional. We repeated the benefits-related questions for each class of the summer school to reveal possible differences among the knowledge areas covered.

# 5 RESULTS

## 5.1 Pre- and Post-Event Survey

Here we report on how the students' responses to the same set of questions, where applicable, evolved from pre- to post-survey. We speculate on how any differences can be attributed to their training.

*Experience* (pre-survey only). In the pre-event survey, we asked the students about their prior experience and knowledge. We did not observe any significant differences in the sample, which was expected since they were all recent graduates or about to start their senior year. All of them had less than a year's worth of full-time-equivalent software development experience based on their course projects and internships. All had internships, and the majority (12 out of 16) characterized their internship experience as "fairly" or "highly" beneficial and expected it to be sufficient for their first full-time job. Only one student expressed doubt about the value of their internship experience. We did not observe notable differences in different knowledge areas.

*Aspirations*. The students' short- and long-term career aspirations showed a shift from pre- to post-survey, suggesting that the training received had an impact on how they perceived their careers. We classified the reported job roles into 12 categories. Development- and management-related roles were equally prevalent and appeared most frequently in the top-three short-term goals in both pre- and post-survey. The popularity of both increased from pre- to post-survey by over 35%. However, the popularity of architecture/design-related roles showed the most dramatic increase, doubling in frequency from pre- to post-survey, in terms of their mention in a top-three short-term spot. When we look at the long-term goals most frequently mentioned in the top-three spots, development-related roles tied with quality-assurance related roles. Management and development-roles registered the most dramatic increase in long-term popularity, at least doubling from pre- to post-survey. Simultaneously, management-related roles by a large margin replaced quality-assurance roles in post-survey to become the runner-up, tied with architecture/design-related roles. The patterns are slightly different if we consider only the top-ranked goal: management-related roles replaced development-related roles as the top-ranked short-term goal in the post-survey. Management-related roles dominated the top spot as the long-term goal in both pre- and post-survey, but increased in frequency. Design/architecture-related roles replaced development-related roles as the runner-up long-term goal in the post-survey. *Thus, while the training raised awareness of development-related roles among the students, it also dramatically biased them toward higher-level roles in management and architecture both in the short and the long term.*

*Software Engineering Activities*. In an open-ended question, we asked the respondents to list a software engineer's day-to-day activities of which they were aware. We grouped the answers from the pre- and post-survey into 38 distinct activity categories and analyzed how the categories shifted. In the post-survey, the students introduced 15 totally new categories, and there was a 65% increase in the number of activities listed from 69 in the pre-survey to 114 in the post-survey. Some categories were specialized to specific activities in the post-survey ("meeting" to "daily meeting", "standup" or "daily scrum"; "test" to "unit test"; and "fix bug" to "debug"), without introducing new categories. We then grouped categories into higher-level themes to gauge increase or decrease in awareness at a more meta level. Table 2 summarizes the theme-level results. One theme ("Commit + Build + Deploy") was entirely new, which suggested that this theme had not been on the students' radar before the training. *We observed a marked increase in awareness in five themes, presumably due to their emphasis in the program.* A marked decrease in awareness was observed in two themes, one of which (requirements engineering and analysis) was not covered in the program and the other of which (next to last row in Table 2) was covered (under *Software Project Management* and *Application Lifecycle Management*). Interestingly, this latter theme relates to the most-aspired-to long-term career goal: a management-related role. Perhaps the students, by omitting the related activities, were subconsciously focusing on short-term goals rather than on longer-term ones that required more experience than they had.

**Table 2. Software engineering activities: shifts in activities reported at the theme-level**

| Theme | Pre-S. Freq. | Post-S. Freq. |
|---|---|---|
| Commit + Build + Deploy | 0 | 8 |
| Test + Unit Test + QA + Review + Code Review | 10 | 22 |
| Fix Bug + Debug | 1 | 6 |
| Meet + Email + Communicate + Collaborate | 11 | 19 |
| Code | 10 | 15 |
| Break + Food + Physical Activity | 1 | 6 |
| Document | 3 | 5 |
| Learn + Research | 3 | 5 |
| Manage + Risk + Conflict + Resolve + Plan + Prioritize + Lead | 11 | 6 |
| Requirements + Analyze | 5 | 1 |

*Software Engineering Concepts*. We asked the students to self-assess their familiarity with a wide range of 39 software engineering concepts covering engineering practices, development processes, project artifacts, design principles, and product artifacts. We summed the Likert-scale scores for each student and compared their pre- and post-survey totals. The mean percentage increase in the total score was 28%. One student's total score markedly decreased from pre- to post-survey, and we attribute it to the student's initial overestimation of his/her familiarity with the given concepts (and subsequent realization of this overestimation). The differences in scores were normally distributed (confirmed with Shapiro-Wilk test and QQ plot), and significant according to the paired t-test with a one-tailed *p*-value of 0.00067. The non-parametric Wilcoxon signed-rank test [18] for paired samples was also significant with a *p*-value of 0.0015. Cohen's *d* for the t-test was 0.98, suggesting a large effect size [19]. *We conclude that training resulted in a significant increase in the students' perception of their familiarity with core software engineering concepts.*

*Specific Topics Covered in Classes.* Finally, for each of the classes covered in the program, we asked the students to self-assess their knowledge and experience in two to four topics related to that class. Again, we aggregated the Likert-scale scores for pre- and post-survey responses. For each class, we performed a paired (repeated measures) analysis to gauge whether the class made a difference. We also aggregated the results over all classes to evaluate the whole program in the same manner. The results are summarized in Table 3. We evaluated the normality of the samples with the Shapiro-Wilk test and applied the one-tailed paired t-test if normality could be confirmed at an alpha-level of 5%. Otherwise, we applied the non-parametric one-tailed Wilcoxon signed-rank (WSR) test for paired samples [18]. We used one-tailed versions of both tests because we naturally expected the scores to improve for each class and overall, barring a systematic self-overestimation in the pre-survey. We set the alpha level at 5%. We measured effect size using Cohen's *d* for the t-test and correlation coefficient (*r*) for the signed-rank test. For Cohen's *d*, following common practice, we interpreted effect sizes exceeding .5 as medium (M) and exceeding .8 as large (L). For the correlation coefficient, we interpreted effect sizes (absolute value) exceeding .3 as medium (M) and exceeding .5 as large (L). Otherwise the effect size was considered small (S) [19].

**Table 3. Topics covered: class-level and overall improvements in self-assessed knowledge and expertise.**

| Class | # of Topics | Norm? | Test used | Sig? | Effect Size |
|---|---|---|---|---|---|
| Software/System Architecture | 3 | Yes | t | Yes | 1.3 (L) |
| Agile Software Development | 2 | Yes | t | Yes | .97 (L) |
| Web Development and JavaScript | 4 | Yes | t | Yes | .47 (S) |
| Application Lifecycle Management | 3 | No | WSR | Yes | .76 (L) |
| Software Verification & Validation | 2 | No | WSR | Yes | .47 (M) |
| Software Craftsmanship | 4 | Yes | t | Yes | .97 (L) |
| Object Orientated Programming with Java | 3 | Yes | t | No | N/A |
| Software Project Management | 2 | No | WSR | Yes | .54 (L) |
| **Overall** | **23** | **Yes** | **t** | **Yes** | **1.47 (L)** |

All tests, except for topics related to the *Object-Oriented Programming* class, were significant with *p*-values much lower than the required alpha level. Effect sizes for all significant tests were large, except for two classes that had a small and a medium effect size.

The *Object-Oriented Programming* class covered core knowledge with which the students were likely to have been most familiar. The insignificant result may thus be explained by its apparent limited added value. In the follow-up survey, this class was one of the knowledge areas that were deemed most useful post-graduation. Students may not have fully realized how much they took away from this class at the time. The importance of the topic could have dawned on them after the fact, when they saw it applied on a daily basis in their jobs.

A similar explanation might apply to *JavaScript Programming and Web Development*, which had a small effect size. However, this class incidentally received the lowest scores in the instruction evaluation, which might have been a more influential factor.

*Overall, we conclude that the summer school had a significant impact in exposing the students to the topics relevant to the organization, as subjectively reported by the students themselves.*

### 5.2 Post-Event Survey Impressions

For a more qualitative impression, we provide selected quotes excerpted from the open-ended comments (the quotes are translated by the authors).

"I developed a better understanding of a lot of theoretical courses covered at my school, with examples from the software industry. We also had the opportunity to apply them. I think this training was very useful. It was very systematically prepared."

"It was easy to interact with the instructors. It was great to have such highly-experienced people to spend time with us."

"I learned a lot of concepts about software engineering that I had never known of or heard about. In addition, the summer school gave me an idea about work life and environment. It was really insightful and helpful from all angles."

### 5.3 Exams

Although we could not administer a pre-event exam for comparison purposes, the students' uptake of the materials covered was objectively and rigorously evaluated by separate post-event exams for each class. All students with a score of 60 or higher received a certificate, and the top-ranking student received a special prize from the company (a new tablet computer). The average overall score was 75% with the lowest score being 60% and the highest being 89%. Exam grades for each student and class are given in Figure 6.

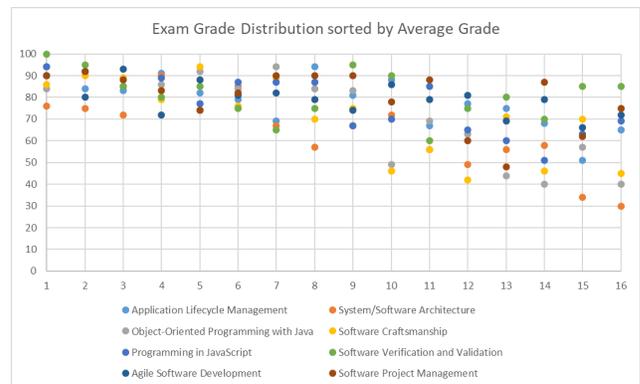

**Figure 6: Exam grades according to knowledge area (class) sorted by average overall grade.**

## 5.4 Retrospective

Following thematic grouping, the *Mad-Sad-Glad* exercise produced 13 items under the *Glad* category, 14 items under the *Sad* category, and 11 items under the *Mad* category. During the plenary discussion on the top-voted items, four items were identified as the most pressing in the *Mad* and *Sad* categories.

*Length and intensity (Mad):* The students indicated that the duration of the summer school was too short for the material covered and the program was too intense. They suggested to either extend the duration and spread the material covered or eliminate some classes. The consensus was that the summer class could be extended to three weeks. *Lodging (Mad):* The company did not provide accommodations for out-of-town students and expected them to make their own arrangements. The students indicated that this was a problem and it stretched their budgets. Another lodging-related suggestion was to have all the students stay at the same place to maximize their time together. *Orientation (Sad):* In the first day, there was a one-hour company presentation followed by a short company tour. The students found this orientation to be too short. They indicated that they would have liked to see the office environments of the engineers and know more about the company. *Timing (Sad):* Since universities have different academic calendars, the students indicated that it would be better if the company could repeat the summer school at two different times or offer it during the winter break as well.

These points were later discussed internally by the organizers to improve the future editions.

## 5.5 Evaluation Survey

The evaluation survey contained both five-point Likert-scale questions (higher scores indicating more positive impressions) and open-ended comments. The questions covered organization, content, and instructor delivery. The average score for the organization-related questions ranged from 3 to 5 with an overall average of 4.47. Low-scored questions and open-ended comments re-confirmed the organizational issues identified during the retrospective. Additional issues with application and registration process were also brought up. These are further discussed in Section 6.4 under lessons learned. On the positive side, the students indicated that the social dinner held on the first day was very beneficial for them to connect with each other and the instructors.

Average instructor evaluation scores ranged from 4 to 4.9. We did not notice anything remarkable in the answers, and could not single out particular issues to address since the lowest score was already on the high end. These questions were sectioned according to class and included content-related questions specific to each class. The results, including open-ended comments, were anonymized and shared with the individual instructors.

## 5.6 Hiring Outcomes and Retention

Out of 16 the students who participated in the summer school, ten were invited for an interview based on their exam performance and on the instructors' feedback. Eight of the ten interviewed students were hired after the interview process. Seven students were hired as software development engineers and one student was hired as a business analyst. As of the writing of this paper, all of the new hires were still with the company. Their performance evaluations, as informally reported by their first-line managers to the summer school organizers, indicated a high level of satisfaction with these employees.

## 5.7 Follow-up Survey

The follow-up survey was conducted over a year after the summer school. All of the 16 participants responded to it. The results were encouraging. 15 out of the 16 students indicated that they would definitely recommend the summer school to their friends. One student would recommend it with some reservations. We also probed the usefulness of the school on a per-class basis. The survey results for each class is summarized in Table 4.

**Table 4: Follow-up survey results.**

| To what extent were the skills learned in this class useful to you in the work environment? | Not Useful | Somewhat Useful | Very Useful |
|---|---|---|---|
| Software Craftsmanship | 1 | 5 | 10 |
| Software/System Architecture | 4 | 6 | 6 |
| Agile Software Development | 2 | 3 | 11 |
| Web Development and JavaScript | 3 | 6 | 7 |
| Application Lifecycle Management | 1 | 6 | 9 |
| Software Verification and Validation | 2 | 7 | 7 |
| Object-Oriented Programming | 0 | 5 | 11 |
| Software Project Management | 2 | 10 | 4 |
| Summer School in General | 0 | 3 | 13 |

According to the responses, *Agile Software Development* and *Object-Oriented Programming* were deemed most helpful, whereas *Software Project Management* was deemed least helpful. A possible explanation for the apparent, relative lower impact of this latter knowledge area early in the students' careers could be attributed to the advanced nature of the topic: project management is a skill that they are more likely to need later in their careers, and none of the participants assumed a project management responsibility during their first year.

In response to the open-ended question about their experiences at the summer school, one participants indicated: *"Within the last year, I have had the chance to experience many of the lessons I had learned at the summer school. The summer school gave me theoretical information applicable in real-life situations… it has made a great contribution."* We did not receive any negative open-ended comments.

## 6 DISCUSSION

### 6.1 Benefits to the Organization

Overall the summer school was beneficial to the company. The company collected many résumés during the selection process. These résumés were shared with the HR department for potential future hires. The company hired eight graduates, a 50% conversion rate, of the summer school, with a 100% retention rate one year after employment. For these students, the summer school also reduced onboarding-related training they would otherwise have received, speeding up their integration to their jobs.

An important side benefit of the summer school was the elevation of the company's image among university students as a cutting-edge software organization that cares about employee development. We believe the summer school program increased the visibility of the company in academia as a preferred employer. The feedback from students provided via multiple surveys and the retrospective were overwhelmingly positive. The negative feedback pertained almost exclusively to organizational and logistic issues and was constructive.

The experiences and lessons learned from the summer school were shared with other business units. Because it was deemed successful, the company repeated the process and format with similar success on more specialized topics within the last year, in addition to a rerun of the same offering in Summer 2017. The Summer 2017 offering accepted the same number of students as the 2016 edition (16 students) and resulted in ten new hires (two more than the 2016 edition), increasing the conversion rate from 50% to 63%.

### 6.2 Benefits to the Participants

The participants, whether hired by the company or not, also appear to have received important benefits. In the post-event and follow-up surveys, all the students rated the summer school as beneficial or very beneficial for them, with a majority choosing the highest rating. In their comments, they indicated that taking a class from practitioners was a lot different than taking a class from university professors. The students also indicated that the type of classes in general were different from typical university classes. Through the experience, they believed they had learned new skills that proved useful in the beginning of their professional lives. We recorded significant improvements in their self-assessed knowledge levels for the majority of the knowledge areas covered. In summary, the summer school was a useful compliment to the more theoretical and formal education that the students had received at university. It provided the students with an extra edge to prepare them for their careers.

The students who were not subsequently hired by the company received a completion certificate that they could leverage in their job search. According to the follow-up survey, the experience helped them in their job search with other companies, with five students indicating to have greatly benefited from it (the number excluding the eight students who were hired by Havelsan) and three students indicating to have benefited from it. Six of these students found employment elsewhere as software development engineer, one found employment as a test engineer, and one student decided to pursue a graduate degree.

### 6.3 Economic Considerations

*Instructor costs:* The summer school lasted two weeks. We estimate that per each day of actual class hour, the instructors spent about three hours for preparation. Thus, the total preparation and teaching effort was about two person-months. A large part of the preparation costs was amortized during the Summer 2017 offering by reusing the instructional materials.

*Organization costs:* These costs were incurred by the main organizer and an associate and includes design, planning, approval, coordination, and candidate screening and selection effort. We estimated it to be about one person-month.

*Logistic Costs:* These costs include promotional materials, hard-copy handouts, meals, breaks, and prizes for the students. The budget was approximately USD 5,000.

After converting all the effort to dollars based on the average salary of the instructors and organizers, the total cost of the summer school to the company was around USD 20,000.

The cost of hiring a new software engineer through the normal process includes creating a job description, advertising, administrative screening of résumés by HR, technical screening by hiring and line managers, phone interviews, multi-step face-to-face interviews, and other onboarding costs. The average cost comes to about USD 3,000 per new hire. Given that the summer school produced eight new hires, and some incremental onboarding costs were still accrued for these employees, we estimate that the summer school at worst broke even. This calculation excludes costs amortized over subsequent offerings, extra training the new hires would have received at the beginning of their employment, and additional, intangible benefits.

### 6.4 Lessons Learned and Limitations

*Candidate eligibility.* The inclusion of $3^{rd}$-year students created a problem for these participants. Both $3^{rd}$-year students indicated that they had had difficulty following some classes and obtained lower grades compared to the $4^{th}$-year students. We amended our eligibility criterion the following year to include only $4^{th}$-year students.

*Duration and intensity.* The majority of the students indicated that they would prefer a longer time-frame for the summer school and complained about the intensity. Although this was a valid point, we could not afford to keep the instructors away from their day-to-day responsibilities and active projects for longer without planning their time off well in advance. In the 2017 offering, we maintained the length. However, our plan for the upcoming years is to do more advance planning with the goal of extending the summer school to three weeks.

*Application and registration.* Issues relating to these processes were a major source of complaints according to the evaluation survey and the retrospective. In 2017, we reduced the documentation requirements, and instituted improvements to simplify and streamline application and registration.

*Student logistics.* For the future years, we are looking into partnering with nearby universities (three major universities are located within a 10-mile radius) to lodge students in dorm facilities underutilized during the summer and introduce a shuttle service. Such a setup would result in significant savings for out-of-town students and create more opportunities for them to socialize, all at a small cost to the company. We would still use the company premises for classes since the participants appreciated the real work environment and proximity to practicing engineers.

*Validity limitations.* Some of our observations rely on self-assessment by the participants. These assessments were triangulated in multiple ways, both qualitatively (retrospective, open-ended questions, informal performance feedback) and quantitatively (rating questions, exams), however they were not conducted in particularly deep ways. We have no reason to suspect strong biases, but we cannot eliminate all biases either. Our analyses are subject to the usual caveats of measurement issues, sample size limitations, construct validity problems due to instrumentation threats and subjective questions, and limited generalizability. They should thus be taken at face value.

## 7  CONCLUSIONS

In 2017, the company repeated the summer school and diversified the offerings by adding new, more specialized curricula in the areas of Data Science, Test Engineering, Cyber Security, and Enterprise Software (SAP). The summer school programs are now an integral part of Havelsan's hiring strategy, with an over 55% overall conversion rate and 100% retention rate so far. Some classes of the summer school curriculum are also now being used during engineering orientation for new hires who do not attend the summer school.

In the future, the company will keep track of performance and turnover of the employees in the software engineering track hired through the regular process versus those hired through the summer school process to better understand the advantages and disadvantages of each.

After repeated successes, the company strongly believes in this pre-hiring training strategy. The benefits have been both tangible (new hires) and intangible (improved image). The cost-benefit analysis indicates that it is economically viable in Havelsan's organizational context, and likely saves money. We also have evidence of multiple benefits to the student community. Although we cannot generalize the results to other organizational contexts easily, we hope that our experience and results will inspire others.

## ACKNOWLEDGMENTS

The authors would like to thank Havelsan management and summer school instructors.